\documentclass[12pt,preprint2]{emulateapj}

\slugcomment{Accepted for publication in ApJ}

\shorttitle{The Two-Point Correlation of 2QZ Quasars and 2SLAQ LRGs}
\shortauthors{Norman et al.}

\begin{document}


\title{The Two-Point Correlation of 2QZ Quasars and 2SLAQ LRGs: From a Quasar Fueling Perspective}


\author{Dara J. Norman\altaffilmark{1}, Roberto De Propris\altaffilmark{2}, and Nicholas P. Ross\altaffilmark{3,4} }
\altaffiltext{1}{National Optical Astronomy Observatories, 950 N. Cherry Ave, Tucson, AZ 85719 ;dnorman@noao.edu}
\altaffiltext{2}{Cerro Tololo InterAmerican Observatory, Casilla 603, La Serena, Chile}

\altaffiltext{3}{Physics Department, Durham University, South Road, Durham, DH1 3LE, U.K.}
\altaffiltext{4}{Department of Astronomy and Astrophysics, The Pennsylvania State University, 525 Davey Laboratory, University Park, PA 16802, U.S.A.
}




\begin{abstract}
Public data from the 2dF quasar survey (2QZ) and 2dF/SDSS LRG \& QSO (2SLAQ), with their vast reservoirs of spectroscopically located and identified sources, afford us the chance to more accurately study their real space correlations in the hopes of identifying the physical processes that trigger quasar activity.  We have used these two public databases to measure the projected cross correlation, $\omega_p$, between quasars and luminous red galaxies.   We find the projected two-point correlation to have a fitted clustering radius of $r_0,  = 5.3 \pm 0.6 $ and a slope, $\gamma =1.83 \pm 0.42 $ on scales from 0.7-27$h^{-1}$Mpc. 
 
We attempt to understand this strong correlation by separating the LRG sample into 2 populations of blue and red galaxies. We measure at the cross correlation with each population.  We find that these quasars have a stronger correlation amplitude with the bluer, more recently starforming population in our sample than the redder passively evolving population, which has a correlation that is much more noisy and seems to flatten on scales $< 5h^{-1}$Mpc. We compare this result to published work on hierarchical models. The stronger correlation of bright quasars with LRGs that have undergone a recent burst of starformation suggests that the physical mechanisms that produce both activities are related and that minor mergers or tidal effects may be important triggers of bright quasar activity and/or that bright quasars are less highly biased than faint quasars. 

\end{abstract}


\keywords{galaxies:general - large-scale structure of the universe - quasars:general}



\section{Introduction}

The connection between quasars and their host galaxy evolution is notably demonstrated by the tight correlation of black hole (BH) mass to host bulge mass, the M$_{BH}$-$\sigma$ relation (e.g.,Gebhardt et al. 2000, Ferrarese \& Merritt 2000).  A question that emerges from this observation is 'how does the BH at the center of a galaxy know what mass bulge it is sitting in?'  A reasonable answer is that both of these masses grow together, and that the growth of both is triggered by the same mechanism.   Since AGN are believed to be the sites where BHs are actively acquiring mass, we are motivated to look to these phenomena for clues to the physical mechanism that drives this correlation.  

A way to explain this connection of BH and bulge mass is to invoke major mergers of galaxies.  Models of hierarchical growth of structures and galaxy evolution have suggested that quasars are fueled by galaxy major mergers (e.g. Haehnelt \& Kauffmann 2000). During these events, gas is driven into the center of the galaxy, feeding the massive black hole within.  Kauffmann \& Haehnelt (2002) use such models to show that the quasar luminosity function can qualitatively be reproduced, albeit with a very simple prescription for fueling and time-scales.
In addition, recent models use the consequence of quasar/AGN phenomena during galaxy mergers to alleviate the long standing problem of
an over-production of high luminosity, blue galaxies while under-predicting the numbers of Milky Way-like galaxies in
 galaxy evolution simulations (e.g.,Croton et al. 2006; Di Matteo et al. 2005).  By providing a heating source for cool gas, quasars/AGN added to these models are able to halt the build up of starformation in their hosts, arresting the growth of these luminous blue galaxies. The suggestion of a further connection between quasar/AGN activity and galaxy evolution is encouraging and leads to the expectation that quasars and AGN are found in overdense regions where the merger fraction is expected to be high. However, there is also evidence that mergers operate more efficiently in overdense regions where the galaxy densities are not too high and so the encounter velocities are lower, like in galaxy groups or at the edges of rich clusters (e.g. Gnedin 2003).  These are regions less likely to be occupied by red elliptical galaxies, more numerous toward the centers of rich clusters.
Although in the most dense regions of clusters merging fractions are expected to be lower, the cumulative effect of weaker interactions like minor mergers or harassment may be important triggers of active nuclei. 

Recent large surveys like Sloan Digital Sky Survey (SDSS), 2dF, 2QZ and 2df/SDSS LRG and QSO (2SLAQ) provide the opportunity to obtain good statistics on the environments of  luminous and dim AGN.  However results of some recent studies on the close environments of AGN provide a mixed picture.
Kauffmann et al. (2003) use an SDSS sample to show that the hosts of narrow line ($z<0.3$) are predominately massive, early-type galaxies (which are generally found in rich environments).  However, for the most luminous of these AGN, there are a high fraction in low density environments (Kauffmann et al. 2004; Coldwell et al. 2002).  Furthermore, Miller et al. (2003)  find that the fraction of galaxies that host AGN ($z<0.1, M_r=-20$) is independent of environment. While, Li et al. (2006) find that narrow-line AGN are more weakly clustered than a  matched sample of SDSS galaxies ($0.01<z<0.3, -23< M_r<-17$).

On the more luminous end of the active nuclei range, Serber (2006) finds that luminous quasars ($z <0.4$, $M_i < -22$) cluster like $L_*$ galaxies for scales $< 1$Mpc. At high z, Coil et al. (2007) report that the SDSS quasars  ($0.7< z <1.4$, $-25<M_B<-22$) that overlap with the DEEP2 survey are found to cluster more like blue galaxies than red. 


  Here we explore one simple test of the major merger models of quasar triggering; Does the spatial distribution of quasars and LRGs support the picture where quasars eventually become LRGs?
Our method is to measure the real spatial cross correlation of 2QZ quasars from the 2QZ survey with galaxies targeted by the 2SLAQ.  The two main advantages of these large surveys in measuring correlation functions are 1) the large number of quasars and galaxies available in the datasets and 2) the availability of spectroscopic redshifts for a large fraction ($> 90\%$) of the sources in the survey volume.
Other studies (e.g. Brown et al. 2000, Padmanabhan et al. 2008) rely on statistical photometric redshifts to determine the association of galaxies with quasars.    

In the next section we describe the data sets used for this analysis. We present our calculations of the quasar-galaxy correlation function in section 3. A discussion of the findings and analysis is presented in Section 4 \& 5 and in Section 6 we discuss our results and we give some concluding remarks.  Throughout this paper we use cosmological parameters $\Omega_m = 0.3$, $\Omega_{\Lambda} = 0.7$ and $H_0 = 70$ km s$^{-1}$ Mpc$^{-1}$.

\begin{figure}
\plotone{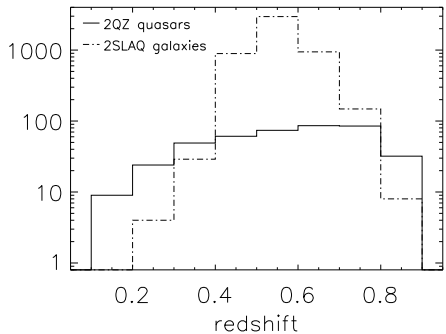}
\caption{Histogram of redshifts for the 420 2QZ quasars (solid) and
 4985 2SLAQ galaxies (dash-dot) used in this 
study.   The median redshifts for each sample are 0.59 and 0.55, respectively. \label{fig1}}
\end{figure}

\begin{figure}
\plotone{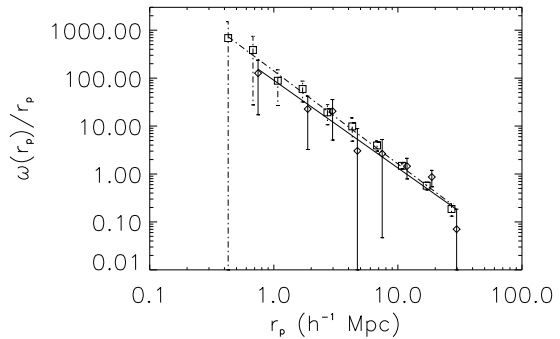}
\caption{The figure shows the projected two point correlation function for 2QZ quasars with 2SLAQ LRGs
plotted as diamonds. 1$\sigma$ 'Jackknife' errors are also shown.  The best fit function of the form in Equation 4
is overplot (solid) for scales 0.7-27$h^{-1}$Mpc. Parameter values are $r_0=5.3 \pm 0.6 h^{-1}$Mpc and $\gamma=1.83 \pm 0.42$. The abscissa positions have been slightly offset for clarity.
 We also show our measured LRG-LRG autocorrelation function
(squares) with the functional fit of $r_0= 7.1^{+0.3}_{-0.4}h^{-1}$ Mpc and
$\gamma = 1.98^{+0.07}_{-0.06}$ (dash-dot).  \label{fig2}}
\end{figure}

\section{The Data}


\subsection{The 2SLAQ Luminous Red Galaxy Sample}
Our LRG sample is taken from the 2SLAQ survey described in Cannon et al. 2006 (hereafter C06).  The public data set (2SLAQ LRG Redshift Catalogue, version 5 available at www.2slaq.info) contains nearly 15000 LRGs with magnitudes brighter than $i=19.8$ along with their measured spectroscopic redshifts. We select only those regions from the northern stripe that overlap the 2QZ sample and do not use parts of the survey where the completeness is low (i.e., fields a01 and a02).  We select galaxies that cover the redshift range from 0.2 to 0.8 with a median redshift of 0.52. Figure 1 shows a histogram of galaxy redshifts. Our sample is selected from the 'top' priority sample described in C06 (Sample 8, selected with priority parameter=8).  These are sources that fall in the most conservative $r-i, g-r$ color cut of the 2SLAQ survey. We further selected only those sources with good spectral quality ($Q \ge 3$).The overall measured redshift completeness for the 2SLAQ Sample 8 (that is,the percentage of the sample with good enough spectra taken to provide reliable redshifts) is 91.4\% (C06, Ross et al. 2007).   The color and magnitude restrictions ensure that our full galaxy sample is dominated by the most massive and luminous intermediate redshift galaxies, these are expected to be galaxies with a passively-evolving stellar population (see Eisenstein et al. 2001).  
Our final galaxy sample contains 4985 LRGS.

\subsection{The 2dF Quasar Sample}
Although the SLAQ survey also includes quasars, we choose our quasar sample from the 2dF QSO Redshift Survey (2QZ, Croom et al. 2004).  The primary reason for this is to avoid sampling problems resulting from survey priorities, fiber placement, etc. within the single survey.   For the 2SLAQ sample, quasars are assigned lower priority in the placement of spectral fibers.  Although there is overlap of the fields that would allow lower priority sources to be picked up in subsequent observations, the subtle vagaries of target selection and fiber placement will lead to a bias in the measurement of the two point correlation function, particularly at smaller scales.  In order to avoid having to make significant corrections for this 'crosstalk' between quasars and galaxies that may vary from field to field, we instead use the overlapping 2QZ survey. This selection limits us to the study of more luminous quasars since the 2QZ targets are brighter than the 2SLAQ survey targets. Although there is no crosstalk between our two samples, there are still issues of fiber placement in both samples, thus we are limited to measuring correlation functions on intermediate scales between about 0.7 - 30 Mpc.  

The 2QZ survey is a flux limited, optically selected quasar survey conducted within the 2df galaxy redshift survey. The survey includes 23,338 quasars with $b_j$ magnitude range of $18.25< b_j <20.85$. Quasar candidates were selected with U-excess in the $U-b_j, b_j-R$ color plane and confirmed with spectra of emission lines. We select QSOs that overlap the non-contiguous region of the 2SLAQ survey within the galaxy redshift range of  $0.1 < z < 0.8$.  Figure 1 shows the redshift distribution of our final quasar sample.  Within this redshift range, we have selected luminous quasars with $-21 > M_b{_j} > -24$ (Croom et al. 2004).
From the public database (www.2dfquasar.org/Spec\_Cat/), we select only those sources classified as 'QSO' that have quality flags of 11, indicating that the spectra are of high quality and redshift measurements are secure. Our final sample contains 420 quasars.




\section{Measuring the Cross Correlation of Quasars and LRGs}
The cross correlation function is used to measure the clustering of sources above a random distribution described by Poisson statistics.  This correlation function is a measure of the excess probability of finding a source from one sample within a volume element, dV,  at a distance r from a source in a second sample.  Here, we use a standard estimator (Landy \& Szalay 1993) to measure the quasar-LRG cross correlation function, $\xi$, where we measure the observed number of LRGs around each quasar as a function of distance and divide by the expected number of galaxies from a random distribution.     

\begin{equation}
\xi(r) =\frac{n_{Random}}{n_{galaxies}}  * \frac{QG(r)}{QR(r)} -1 
\end{equation}
Here QG is the number of quasar-galaxy pairs and QR is the number of quasar-random pairs at a given separation. $n_{random}/n_{galaxies}$ is a normalization factor for the mean number densities in each catalog.                                       

We create random catalogs of LRGs that mimic the spatial and redshift distribution of our galaxy sample. Because of the non-contiguous nature of the 2SLAQ survey in right ascension, in order to properly distribute the random sample galaxies,  we use the same RA positions of the real galaxies. We then randomize the redshift and declination of each mock galaxy.  For these catalogs, each source redshift (declination) is selected at random from the observed redshift (declination) probability distribution.  
We use a density of random sources that is 50 times our galaxy density.

We expect apparent distortions of the clustering pattern ('Fingers of God') due to peculiar motions of galaxies along the line of sight (Peebles, 1980).  To account for these distortions, we measure $\xi$ as a function of both $r_p$, component perpendicular to the line of sight and $\pi$, the component along the line of sight.   The real space correlation $\xi(r)$, without distortions,  can then be recovered by integrating $\xi(r_p,\pi$) along the line of sight to calculate $\omega(r_p)$, the projected correlation function.     
 \begin{equation}
   \omega(r_p)  =2 \int^{\infty}_{0}  \xi(r_p,\pi) d\pi                        
\end{equation}                                           
 It has been shown by Davis and Peebles (1983) that the projected correlation function is related to the real-space correlation function through the equation,

\begin{equation}
   \omega(r_p)  =2 \int^{\infty}_{r_p} rdr \xi(r) (r^2-r^{2}_{p})^{-0.5}          
\end{equation}  
  
In practice, we integrate not to infinity but to some large value, $\pi_{max}$ = 50$h^{-1}$Mpc, which includes almost all correlated pairs and peculiar velocities. We have varied the value of $\pi_{max}$ using both 30$h^{-1}$ and 50$h^{-1}$Mpc and find that our results for 
$r_0$ and $\gamma$ are changed by $<$ 10\%.   Our amplitudes (measured at 1.7$h^{-1}$Mpc) change by as much as 20\%, however this change has no effect on our conclusions since amplitude differences between our two subsamples (see section 5.1) are larger than these values.

The real-space correlation, $\xi(r)$, can then be approximated by a power law.  The projected correlation function is 

\begin{equation}
\omega(r_p)  =r_p \left (\frac{r_p}{r_0} \right )^\gamma  ~\frac{\Gamma(1/2)\Gamma((\gamma-1)/2)}{\Gamma(\gamma/2)]}
\end{equation}

where $\Gamma$ is the Gamma function.  However, real-space correlation function is not expected to be a true power law on large scales ($r_0 > 20h^{-1}$ Mpc). 
Below we present the results of our cross correlation along with power law fits to the function. 

\begin{figure}
\plotone{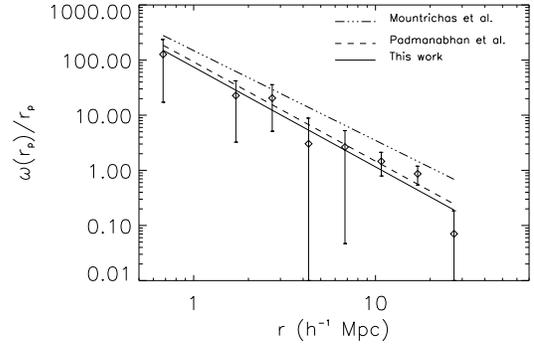}
\caption{In this figure, we compare our measured two point correlation function of 2QZ quasars with the full sample of 2SLAQ LRGs (diamonds) to those of other recent work. Mountrichas et al. (2008) find  $r_0=8.68 \pm 1.0$ and $\gamma=1.63 \pm 0.15$ over scales of 1 to 25 $h^{-1}$Mpc. 
Their selection criteria is somewhat different than ours limiting the redshift range to $0.35 <$ z $< 0.75$ and effectively eliminates more bluer galaxies. There result gives a much larger $r_0$ and and a more shallow slope.
We also plot the correlation function from Padmanabhan et al. (2008) between SDSS quasars and LRGs at $0.2 < $ z $< 0.6$. They measure a correlation function fit by a power law slope of $1.8\pm 0.1$ and $r_0 =6 \pm 0.5 h^{-1}$Mpc over scales of 0.1-20$h^{-1}$Mpc, which compares well to what we find.\label{newrev2comp.eps}}
\end{figure}

\begin{figure}
\epsscale{1.2}
\plottwo{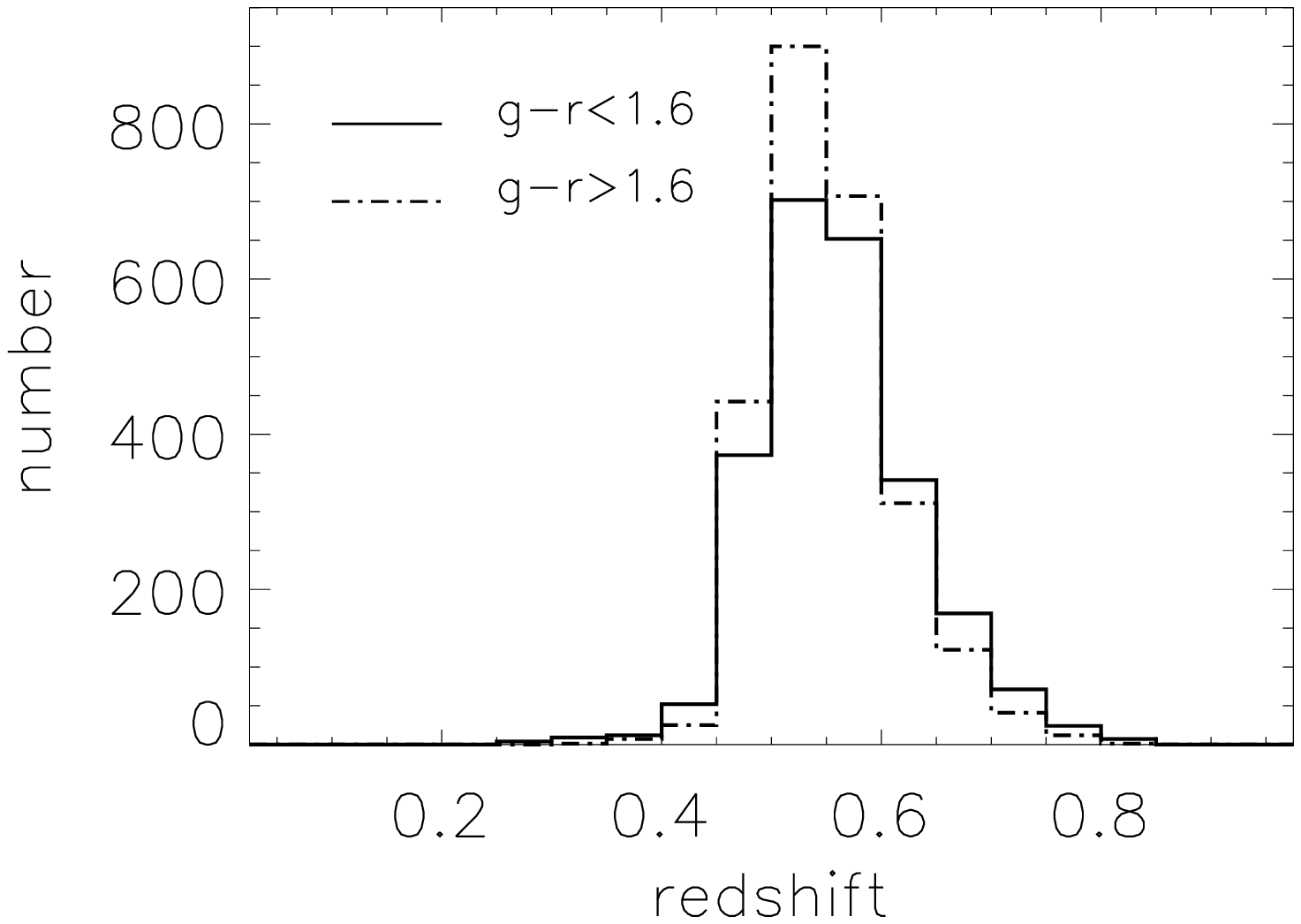}{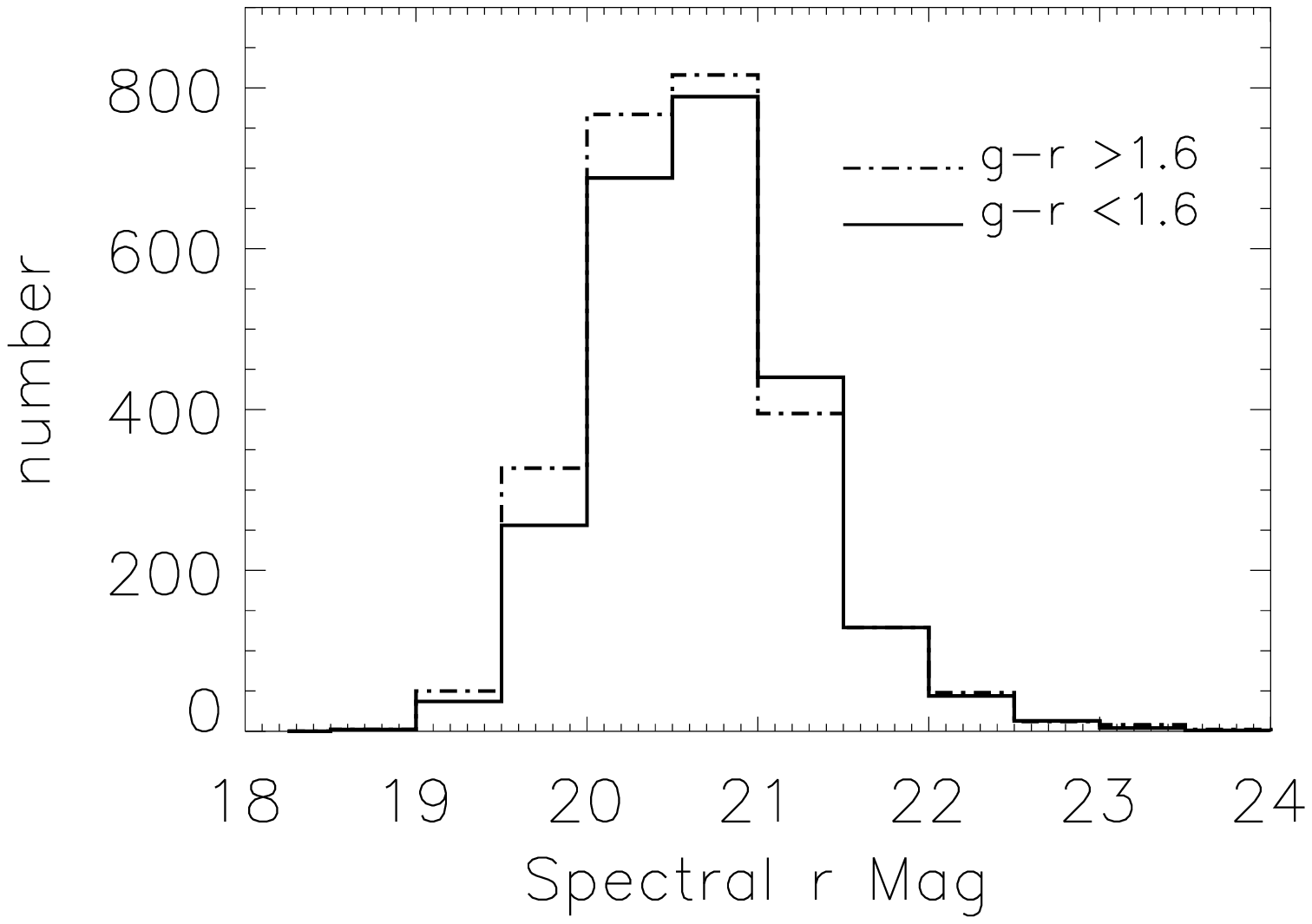}
\caption{The top plot in this  figure shows a histogram of redshifts for the red (2569) and
 blue (2416)  2SLAQ LRG samples partitioned at $g-r=1.6$.  The distributions of the redshifts 
are similar with medians of 0.54 and 0.55, respectively.  The bottom plot is a histogram of r 
magnitudes for the same red and blue 2SLAQ galaxy samples. Again, the distributions are very similar.
 \label{fig3}}
\end{figure}

\section{Quasar-LRG Correlation Results for the Full Sample}
Figure 2 shows the measured projected correlation function between our selected sample of 2QZ quasars and 2SLAQ LRGs.  Although we have used the largest sample of LRGs with spectrally measured redshifts, our ability to measure the correlation on small scales ($<0.5$Mpc) is hampered by the small number of quasar-galaxy pairs in each redshift bin.  Errors have been estimated by using the 'jackknife' method of Scranton et al. (2002), where we divide the sample into 12 approximately equal area regions and calculate the correlation function using combinations of only 11 regions, leaving out the one region for each re-calculation.

\begin{equation}
\sigma_j^2(r_p) = \sum^{N}_{i=1} \frac{DR_i(r_p)}{DR(r_p)} [\omega_i(r_p)-\omega(r_p)]^2
\end{equation}

$DR_i$ is the number of data-random pairs for each subsample, $DR$ is the number of pairs in the  full sample, thus the ratio is a normalization that accounts for the fact than the number of random pairs may not be the same for each subfield (Myers et al. 2005). 

We have done a $\chi^2$ fit of the power law model defined in Equation 4 to the correlation over the range of $0.7-27 h^{-1}$Mpc and find a clustering radius, $r_0 = 5.3 \pm 0.6$ and slope, $\gamma =1.83 \pm 0.42$.  
In Figure 3, we compare our results to Mountrichas, et al. (2008) who have also recently measured the cross correlation of 2QZ quasars and 2SLAQ LRGs.  They find  $r_0=8.68 \pm 1.0$ and $\gamma=1.63 \pm 0.15$ over scales of 1 to 25 $h^{-1}$Mpc. 
They have used a different selection criteria to ours limiting the redshift range to $0.35 <$ z $< 0.75$.  This selection effectively 
eliminates more bluer galaxies than red from the LRG sample at both ends of the redshift range (see figure 3).  Thus they have selected a more intrinsically bright sub-sample of galaxies, irrespective of color (see Figure 4).  If we use a selection more similar to theirs, we find a somewhat larger $r_0$, but we do not match their value.


We have also measured the autocorrelation function for the sub-sample of 2SLAQ galaxies used in our selected sample.  
The auto correlation function is also plotted in Figure 2.  A fit to the measured function gives $r_0=7.1^{+0.3}_{-0.4} h^{-1}$Mpc and $\gamma=1.98^{+0.07}_{-0.06}$ for scales of 0.7 to 47$h^{-1}$Mpc.  This result closely matches that of Ross et al.(2007)  who find $r_0=7.30 \pm 0.34 h^{-1}$Mpc and $\gamma=1.83 \pm 0.05$ over similar scales from 0.4 to 50$h^{-1}$Mpc from a measure of the projected correlation function of the full 2SLAQ galaxy sample.  We therefore conclude that galaxy distribution properties or our selected galaxy sample are representative of the full 2SLAQ LRG sample.

We confirm that the quasar-LRG correlation has a somewhat lower amplitude than the LRG-LRG correlation. In particular at scales $< 3h^{-1}$Mpc, the correlation function seems to be more shallow indicating that a single sloped power law may not be the best fit.  We conclude that our luminous quasars are are strongly correlated with these LRG galaxies, although they seem to be less strongly correlated at scales  $< 3h^{-1}$Mpc.  To try and make some sense of what this means for quasar triggering scenarios, we take a look at the makeup of our LRG sample in the sections below.

\begin{figure}
\epsscale{1}
\plotone{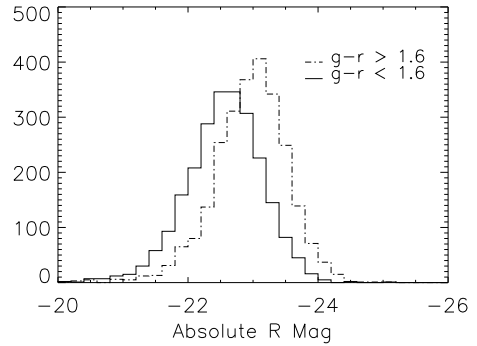}
\caption{The plot in this  figure shows histograms of r-band absolute magnitudes for the red and blue LRG sample partitioned at $g-r=1.6$.  Shown here is that the red sample of galaxies is, on average, more luminous than the blue sample.  We interpret this to mean that the red sample is more massive than the blue sample.
 \label{newf5}}
\end{figure}

\section{Understanding the LRG Sample}
Roseboom et al. (2006)  have looked in detail at the galactic content of the 2SLAQ LRG sample. Based on the equivalent widths of H$\delta$ and [O$II$], they identify four spectral types of galaxies represented.   The majority of the galaxies in the sample (81\%) are found to be passively evolving, based on the absence of absorption or emission lines. 
Indeed, an inspection of the original selection of the 2SLAQ LRG targets from Eisenstein et al. (2001) (also see Roseboom et al. 2006), shows that the evolutionary track model for a passively evolving early-type galaxy runs nearly vertical in color-color space at $g-r\sim 1.7$ through the $r-i$ range where these galaxies have been selected. 
13\% of the galaxies show [O II] emission without H$\delta$ absorption, while 4\% show H$\delta$ absorption without significant [O II] and 2\% have both H$\delta$ absorption and [O$II$] emission.  The presence of emission and absorption lines are believed to indicate ongoing activity in these galaxies; H$\delta$ absorption in 6\% of these galaxies indicates recent starformation which makes these galaxies appear bluer in the LRG sample.  Roseboom et al. (2006) suggest that this starformation is unlikely to be continuous since evolutionary tracks for such a scenario barely skim the color cut imposed by the 2SLAQ survey.  Thus, they conclude that the starformation in these galaxies is the result of 'instantaneous' recent bursts.  

The sources in the 2SLAQ LRG sample that are classified as emission line galaxies ($\sim 13\%$) may be the result of either ongoing starformation or AGN activity.  Roseboom et al. 2006 were unable to distinguish between these because of the lack of spectral coverage of the diagnostic lines.  Therefore, observationally $\sim 20\%$ of the 2SLAQ LRG sample are seen to have recent active starformation or perhaps even ongoing AGN activity. 

Wake et al. (2006) demonstrate that at the bluer end of the 2SLAQ sample, magnitude errors push less intrinsically luminous galaxies into the selected LRG sample. Their analysis of the photometric errors shows this to be a 22\% effect in the full 2SLAQ sample.  They also show (in Figure 1 of their paper) that for the redshift range covered, evolutionary models that include starformation are much better fits to the colors of galaxies blueward of $g-r=1.6$.  Thus our LRG sample seems to be made up of 2 different galaxy populations that have different SEDs and have had very different evolutionary histories. One less luminous blue population with on going activity and a second red, more luminous population that is passively evolving.  In the next section, we examine the correlation of 2QZ quasars with these two different galaxy populations.  

\begin{figure}
\epsscale{1}
\plotone{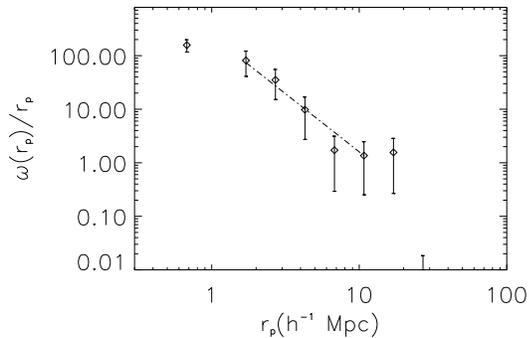}
\caption{Shown is the projected two-point correlation function for 2QZ quasars and blue galaxy sample with 1 $\sigma$ errors.  The best fit to Equation 4
gives a $r_0 =7.7 \pm 0.8$ and $\gamma = 2.17^{+0.40}_{-0.30}$ for scales 1-10$h^{-2}$Mpc. This is a stronger correlation than that for the full LRG sample.  \label{fig6}}
\end{figure}

\begin{figure}
\plotone{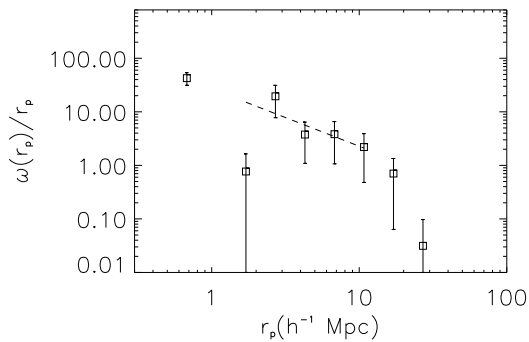}
\caption{The projected two-point correlation function for 2QZ quasars and red LRGs is shown in this plot. The correlation function is noisy despite the larger number of galaxies in the red sample.  The fit for scales of 1-10$h^{-2}$Mpc give a $r_0 = 0.9 \pm 0.1$ and $\gamma = 1.07\pm 0.21$, however the correlation flattens on scales $< 5 $.\label{fig7}}
\end{figure}

\subsection{Quasar-LRG Correlation Results for Red/Blue sub-Samples}
A division of our selected sample at $g-r =1.6$ leaves us with similar numbers of galaxies in the red (2569) and blue (2416) subsamples.  Figure 3a shows a histogram of the redshift ranges for these red and blue subsamples for the top priority 2SLAQ sample and demonstrates that the distributions are quite similar with median redshifts of 0.54 and 0.55, respectively, suggesting that differences in the quasar-galaxy correlation function for each of these samples is not due to redshift evolution. A histogram of the r magnitudes for each of the samples is shown in Figure 3b.  Again, these distributions are similar. 

 Although these 2 galaxy sub-samples have similar distributions in redshift and magnitude, we have shown above that they likely have different spectral energy distributions (SEDs) - about half of galaxies in our blue sample are starforming, and possibly disk galaxies. Thus the absolute r-band magnitude distributions are not the same for these populations.   We estimate the absolute magnitude differences in the distributions by assuming an early type galaxy SED for our red population and a late type distribution for our blue galaxies, which reflects the increased star formation in these sources. We determine r-band K-corrections for these populations assuming an average redshift of 0.55.  Figure 4 shows the difference in the derived absolute magnitudes for each sample.   The red sample is shown to have a higher average luminosity than the the blue sample, as shown by Wake et al. (2006). We assume that mass follows light in these galaxies and that this higher luminosity implies, on average, higher masses for our red galaxies. 

As we did for the full LRG sample, we now measure the projected correlation function, $\omega_p$, for our red and blue galaxy subsamples with the full 2QZ quasar sample.  The correlations are plotted in Figures 5 and 6 with their one sigma Poisson errors. 
We use the Poisson errors here because we found that the jackknife errors seem to misrepresent the true error in our correlation function.  The size of the subsamples are not large enough to partition the samples into enough bins for the jackknife estimate to be robust while keeping the number statistics reasonable for each partition.  Other studies (e.g. Ross et al. 2007, da Angela 2008) have shown that at least for small scales ($< 10h^{-1}$Mpc), where bins are independent, the Poisson errors are similar to the jackknife and field-to-field errors. 

The fitted $r_0$ and $\gamma$ are reported in Table 1 along with the reduced $\chi^2$ of the fit.  We find that the correlation function for the quasar-blue galaxy sample on scales 0.7 to 27$h^{-2}$Mpc has a best fit $r_0=7.3 \pm 0.7$ and $\gamma=2.11^{+0.15}_{-0.13}$ with higher amplitude than the cross correlation with the entire galaxy sample although the reduced $\chi^2$ value indicates that correlation function is poorly fit by a single power law over this range.   On scales of $\sim~1-10 h^{-2}$Mpc however the function is well fit by a power law with $r_0=7.7 \pm 0.8  h^{-2}$ and  $\gamma=2.17^{+0.40}_{-0.30}$, an $r_0$ value larger than that measured for the full galaxy sample.

In contrast, the red sample correlation is very noisy, despite the larger numbers of galaxies in the sample. The power law fit over scales  0.7 to 27$h^{-2}$Mpc gives a $r_0=4.9 \pm 0.7$ and $\gamma=1.66 \pm 0.10$ , but once again, these are poor fits to a single power law model. The fit over a narrower range of scales, $\sim 1-10 h^{-2}$Mpc, gives a nearly flat fit with a formal $r_0=0.9 \pm 0.1$ and $\gamma=1.07 \pm 0.21$. The data are very noisy because the correlation signal is weak, particularly on small scales. This is reflected in the red sample errors.

We also measure the auto-correlation of our red and blue LRG samples, shown in Figure 7.  Results of the power law fits are listed in Table 1.  We find that the redder, more massive LRGs, have a higher clustering amplitude than the bluer galaxies. That is, there is a higher probability of finding a red-red galaxy pair than a blue-blue pair, at least at small scales. Thus the redder galaxies are more strongly clustered than the bluer galaxies, although the clustering radius (where the distribution becomes random) is larger for red galaxies.

\section{ Discussion}

From the above results, we conclude that most of the signal in the measured projected two point correlation of bright 2QZ quasars cluster tothe full sample of LRGs on scales $<$ 5 Mpc is the result of a stronger correlation of quasars with the bluer galaxies in the LRG sample.  While there is a measured correlation with the red galaxies, the signal is weak and flattens on scales $<5 h^{-1}$Mpc (see Figure 5).

Our result that 2QZ quasars seem to cluster more strongly with the bluer galaxies than red galaxies is an interesting one.  The expected merger driven evolutionary history of the reddest LRG galaxies suggests a scenario in which these galaxies occupy the highest mass halos where they are built up through major mergers of galaxies in the highest density regions, which trigger a luminous quasar phase, we might have expected to detect a strong correlation between the most luminous red galaxies and our bright quasars, indicating that they occupy dark matter halos of similar mass. 
 The lack of a strong correlation with bright quasars suggests that these quasars are not just found in the highest density regions.  Our finding is qualitatively supported by Mountrichas et al. (2008), who find that luminous quasars may be less highly biased than faint quasars when compared with LRGs, that is, that the more luminous quasars are not only hosted in the most overdense mass peaks. We explore 2 explanations for our measured LRG subsample correlation functions.
\begin{table}
\begin{center}
\caption{Table of fit data for correlation functions.\label{tbl-1}}
\begin{tabular}{ccccc}
      & $r_0$ & $\gamma$ & $X^2_{reduced}$ & scales \\ 
      & h$^{-1}$Mpc &     &               & h$^{-1}$Mpc \\
\tableline
\tableline
\\

LRG-LRG & $7.1^{+0.3}_{-0.4}$ &$ 1.98^{+0.07}_{-0.06}$ & 0.24  & 0.4 - 45  \\ \\

2QZ-LRG & $6.3^{+0.6}_{-0.5}$ & $1.96^{+0.14}_{-0.20}$ & 1.09 & 4.0 - 27   \\ 
2QZ-LRG & $5.3 \pm 0.6 $ & $1.81 \pm 0.42$ & 1.86 & 0.7 - 27   \\ \\

2QZ-LRG$_{blue}$ & $7.3 \pm 0.7$ & $2.11^{+0.15}_{-0.13}$  & 3.41  & 0.7 - 27 \\
2QZ-LRG$_{blue}$ & $7.7 \pm 0.8$ & $2.17^{+0.40}_{-0.30}$  &0.33 & 1 - 10 \\ \\

2QZ-LRG$_{red}$ & $4.9 \pm 0.7$ & $1.66 \pm 0.10$ & 2.09 & 0.7 - 27  \\
2QZ-LRG$_{red}$ & $0.9 \pm 0.1 $ & $1.07\pm 0.21 $ & 0.9 & 1 - 10  \\ \\


LRG$_{blue}$-LRG$_{blue}$ & $4.9 \pm 0.2$ & $1.77 \pm 0.23$ & 0.68 & 0.7 - 27  \\
LRG$_{red}$-LRG$_{red}$ & $6.5 \pm 0.3$ & $2.00 \pm 0.21 $ & 1.21 & 0.7 - 27  \ 
\\

\end{tabular}
\end{center}
\end{table}

\begin{figure}
\plotone{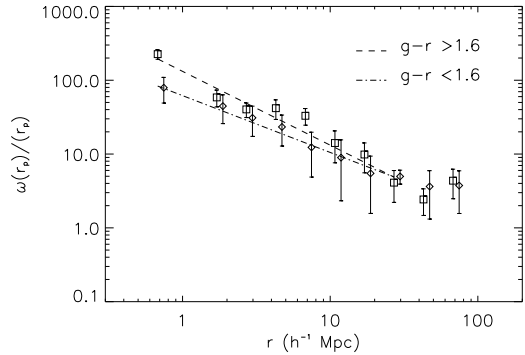}
\caption{The projected auto correlation function for blue and red galaxies are shown along with the fit over scales 0.7-27$h^{-2}$Mpc. The blue galaxy data points have been slightly offset for clarity. Fitted values for $r_0$ and $\gamma$ are listed in table one.  The red galaxies are more strongly clustered particularly on small scales. \label{fig8}}
\end{figure}

We have demonstrated that a large fraction of the blue galaxies in the LRG sample are generally lower luminosity ($\sim L_*$) galaxies  that have undergone some recent burst of starformation and a significant number of them may even host (low luminosity) AGN.  Our strong correlation of quasars with these blue galaxies suggests that it is these galaxies that have space densities similar to our bright quasar sample.
The fact that we exclusively measure a strong correlation of luminous quasars with the population of galaxies that appear to have had recent activity (either an instantaneous burst of starformation or low luminosity AGN) supports models that predict the association of starbursts with AGN activity (e.g., Springel et al. 2005).
 We suggest that environmentally dependent triggering mechanisms for the bursts of starformation in these galaxies and the quasar activity may be the same.   We are unable to say what that mechanism might be; that is whether minor mergers and harassment through tidal interactions might trigger activity or major merger events are contributing.  However, if some of these lower luminosity galaxies do have disk components, they are not expected to be the resultant population of major mergers in hierarchical models.

This suggestion is strengthened by the observation that red galaxies in our sample do not have space densities comparable to or bright quasar distribution.  Our measurement of a turnover in the correlation function on scales $<5h^{-1}$ Mpc between quasars and only the red, more centrally concentrated galaxies reflects the lack of quasar-red LRGs pairs above a random distribution on small scales.  This can be explained if a significant fraction of 2QZ quasars are distributed in regions outside the more centrally clustered red galaxies. This leads us to conclude that the processes responsible for growing LRGs (presumably major mergers) are not those that responsible for triggering most bright quasars at this redshift.

Recent work on the growth of galaxies in hierarchical simulations by Guo and White (2008) lends support to
our conclusions that major mergers may not be the entire story when it comes to triggering luminous quasar. They look at the growth of galaxies through major mergers, minor mergers and starformation for galaxies in a series of binned mass ranges over cosmic time.   They find that the dominant mode of galaxy assembly is a strong function of the stellar mass of the galaxies. For galaxies with lower masses ($M < 8\times10^{10} M_{\sun}$ ), starformation and minor mergers are the dominant modes for galaxy growth.  Whereas for higher mass galaxies major mergers are much more important for growing the galaxies at all redshifts.   This work mirrors what has been found in the 2SLAQ sample, that the lower mass (blue, $L_*$) galaxies are more closely associated with starformation activity, here in the form of bursts, than the more massive (red) galaxies.  
The strong correlation of quasars with these blue galaxies, suggests that minor mergers should be important triggers of activity, even for these bright quasars. 

However, Wake et al. (2008) conclude that there is major merging going on within the LRGs population over redshifts of 0.5 to 0.19. By comparing the autocorrelation function at high and low-z with passive evolution and HOD models, they find that the merger rate over this redshift range is 2.4\% per Gyr. The 2SLAQ LRG sample they use is most similar to our red galaxy sub-sample. Thus for an assumed scenario where major mergers result in quasars, we would expect a strong correlation of red LRGs and quasars. So why is this not what we measure? It is likely that our selection of bright 2QZ quasars does not enable us to measure a correlation because quasars resulting from these mergers are faint due to dust obscuration.  This is consistent with work suggesting that faint quasars are more highly biased than bright quasars. 

We also point out the recent work of Padmanabhan et al. (2008) on correlations of SDSS quasars and LRGs at $0.2 < $ z $< 0.6$. Although they have a larger number of LRGs (chosen similarly to our sample) and quasars in their sample, they do not have the directly measured redshifts for each galaxy, as we have.  They measure a correlation function fit by a power law slope of $1.8\pm 0.1$ and $r_0 =6 \pm 0.5 h^{-1}$Mpc over scales of 0.1-20$h^{-1}$Mpc (see Figure 3).  With their very large numbers of quasars and galaxies, they are able to measure the correlation function to smaller scales than is possible for our sample.  They model their correlation function with halo occupation distribution models where quasar host galaxies populate their dark matter halos preferentially at the centers and with varying fractions of quasar hosts distributed as 'satellite' galaxies within halos. They find that in order to match the correlation function, a significant fraction ($> 25\%$) of quasars must be hosted outside the central halo region in satellite galaxies. 

Coil et al. (2007) have looked at the correlation function of very luminous ($-25.2 < M_{b_{j}} < -22.1$) SDSS quasars with galaxies in the DEEP2 field at about redshift = 1.0.  They find that quasars are found in regions with mean overdensities more similar to blue galaxies than red galaxies. We note however that their blue galaxy sample is, on average, much more blue than our blue sample.   Their selection of blue galaxies are those that have $U-B < 1$ and constitute a more clearly disk dominated population. Because targets for the 2SLAQ survey were chosen to mostly exclude such galaxies, we are unable to compare directly to a similar sample at lower redshift here.   However, although a smaller fraction of our galaxies might be disk dominated, the similarity is in what makes these galaxies blue and that is the increase in starformation over the reddest LRGs.  Like the higher redshift sample, we too find that quasars at z=0.5 are closely associated and cluster like galaxies with enhanced starformation.

\section{Summary and Conclusions}
We have measured the projected two-point clustering of quasars from the 2QZ survey and 2SLAQ LRG galaxies.  We find that $r_0 = 5.3 \pm 0.6 h^{-2}$Mpc  and $\gamma = 1.83 \pm 0.42$ on scales 0.7 to 27 $h^{-1}$Mpc.
A literature search of the galaxy populations that make up this full LRG sample, revealed that our galaxy sample contains two galaxy populations that might have very different evolutionary histories.  Thus we sought insight to our goal of exploring quasar triggering by looking at the clustering signal with each of these galaxy populations.

We find that the clustering signal measured with the full galaxy sample can be attributed primarily to the strong clustering of our quasar sample with the bluer LRGs. Analysis of the spectra of these blue galaxies taken from the literature shows that they are primarily galaxies that have most likely undergone some recent instantaneous burst of starformation or may even harbor low luminosity AGN.  

We further find that the correlation signal with the sample of the most luminous and red LRGs is weak, noisy, and seems to flatten on scales $< 5$ Mpc.  This strong correlation with blue galaxies and flattening with red, more centrally clustered galaxies can be explained by either a significant fraction of quasars being hosted in galaxies outside the cores of clusters or/and  that bright quasars are less highly biased than faint quasars so that our selection criteria excludes most of the faint quasars that are correlated with our red sample of LRGs.

We speculate on what this correlation means for the triggering of luminous quasars.  
The strong correlation between bright quasars and the bluer galaxies in our sample LRG suggests that the triggering of these quasars is related to the starformation in these galaxies at z=0.5. Other studies show that this activity is more likely to be the result of minor mergers or tidal effects between galaxies than major merger events for these lower mass galaxies. Thus minor mergers and/or tidal effects may be important triggers of even bright quasars.

If the spatial distribution of our quasars compared to LRGs is responsible for our observed stronger correlation with bluer galaxies, then this may support the so called ``cosmic downsizing'' scenario (Cowie 1996, Broadhurst et al. 1988), where at higher redshift, quasar activity was more prevalent in high mass galaxies.  By redshift 0.5, this activity has ended, perhaps leaving behind a passively evolving LRG.  The bulk of activity at this lower redshift is now in less luminous, lower mass galaxies. 
However, if the redder LRGs that are still merging lead to quasar ignition, then there may still be some of quasar activity in high mass galaxies at z=0.5, but it is obscured by dust.







\acknowledgments

We would like to thank the 2QZ and 2SLAQ teams for making there data publicly available and easy to use. Thanks to the anonymous referee for very useful comments and suggestions.
DJN would like to acknowledge support from the NSF Astronomy and Astrophysics Postdoctoral Fellowship.

\clearpage


\begin{thebibliography}{}
\bibitem[Brown et al. (2000)]{br00} 
\bibitem[Broadhurst et al. (1988)]{bro88} Broadhurst, T. J., Ellis, R. S., \& Shanks, T. 1988, \mnras, 235, 827 

\bibitem[Cannon et al. (2000)]{can06} Cannon, R., et al. 2006, \mnras, 376, 425
\bibitem[Coil, A. et al. (2007)]{coil07} Coil, A., Hennawi, J., Newman, J., Cooper, M. \& Davis, M. 2007, \apj, 654 115
\bibitem[Coldwell et al. (2002)]{col02} Coldwell, G., Martínez, H. \& Lambas, D. 2002, \mnras, 336, 207
\bibitem[Cowie et al.(1996)]{cow96} Cowie,L., Songaila, A. ,Hu, E. \& Cohen, J. G. 1996,\aj,112, 839
\bibitem[Croom et al.(2005)]{crm05} Croom, S.,et al. 2005,\mnras, 356,415
\bibitem[Croom et al.(2004)]{crm04} Croom, S.,et al. 2004, \mnras, 349, 1397  
\bibitem[Croton et al. 2006]{cro06} Croom et al. 2004, \mnras, 365, 11
\bibitem[da Angela et al.(2006)]{daA06} da Angela, J., et al. 2008, \mnras, 383, 565
\bibitem[Davis \& Peebles (1983)]{dav83} Davis, M. \& Peebles, P.J.E. 1983,\apj, 267, 465
\bibitem[di Matteo et al. (2005)]{dim05} Di Matteo, T., Springel, V., Hernquist, L. 2005, Nature, 433, 604
\bibitem[Eisenstein et al. (2001)]{eis01} Eisenstein et al. 2001, \aj, 122, 2267
\bibitem[Ferrarese \& Merritt 2000]{fer00} Ferrarese, L., Merritt, D., 2000, \apj, 539, 9
\bibitem[Gebhart et al. 2000]{geb00} Gebhardt, K. et al., 2000,\apj, 539, 13
\bibitem[Gnedin 2003]{gne03} Gnedin, O. 2003 \apj, 582, 141
\bibitem[Guo 2008]{goa08} Guo, Q. and White, S.D.M. 2008 \mnras, 384, 2
\bibitem[Haehnelt \& Kauffmann 2000]{hk00} Haehnelt, M.G. \& Kauffmann, G. 2000 \mnras, 318L, 35
\bibitem[Hawkins et al. (2003)]{haw03} Hawkins et al. 2003, \mnras, 346, 78
\bibitem[Kauffmann \& Haehnelt 2002]{kh02} Kauffmann, G. \& Haehnelt, M. G. 2002, \mnras, 332, 529
\bibitem[Kauffmann et al. 2003]{kau03} Kauffmann et al. 2003, \mnras, 346, 1055
\bibitem[Kauffmann et al. 2004]{kau04}  Kauffmann et al. 2004, \mnras, 353, 713 
\bibitem[Landy \& Szalay 1993]{lan93} Landy, S.D. \& Szalay, A. S. 1993, ApJ,412, 64
\bibitem[Li et al.(2006) ]{li06} Li, C., Kauffmann, G., Wang, L., White, S., Heckman, T.,\& Jing, Y. 2006, \mnras, 373, 457
\bibitem[Miller et al. (2003)]{mil03} Miller, C.J., Nichol, R. C., Gómez, P.L., Hopkins, A. M. \& Bernardi, M.  2003, \apj, 597, 142
\bibitem[Mountrichas et al. (2008)]{mou08} Mountrichas, G., Shanks, T., Croom, S. M., Sawangwit, U., Schneider, D., Myers, A. \& Pimbblet, K. 2008 arXiv0801.1816
\bibitem[Myers et al. (2005)]{mye05} Myers, A. D., Outram, P. J., Shanks, T., Boyle, B. J., Croom, S. M., Loaring, N. S., Miller, L., Smith, R. J.2005, \mnras, 359, 741
\bibitem[Padmanabhan et al. 2008]{pad08} Padmanabhan, N., White, M., Norberg, P., \& Porciani 2008 Astroph 0802.2105v2
\bibitem[Peebles (1980)]{pee80} Peebles, P.J.E. 1980, The Large-Scale Structure of the Universe, Princeton University Press, Princeton NJ 
\bibitem[Roseboom et al. (2006)]{rose06} Roseboom I. et al. 2006, \mnras, 373, 349
\bibitem[Ross et al.(2007)]{ross07} Ross, N.P. et al. 2007, \mnras, 381, 573
\bibitem[Scranton et al. 2002]{scr02} Scranton, R. et al. 2002, \apj,579, 48
\bibitem[Serber et al. (2006)]{ser06} Serber, W., Bahcall, N., Ménard, B., Richards, G. 2006, \apj, 643, 68
\bibitem[Springel, et al. (2005)]{spr05} Springel, V., Di Matteo, T., Hernquist, L. 2005, \apj, 620, 79
\bibitem[Wake et al. (2006)]{wak06} Wake et al. 2006, \mnras, 372, 537
\bibitem[Wake et al. (2008)]{wak08} Wake et al. 2008, \mnras, 387, 1045

\end{thebibliography}
\end{document}